\documentclass[12pt]{article}

\usepackage{scicite}

\usepackage{times}

\usepackage{amsmath}
\usepackage{amsfonts}
\usepackage{amssymb}
\usepackage{graphicx}

\usepackage{caption}
\newenvironment{sciabstract}{%
\begin{quote} \bf}
{\end{quote}}

\newcommand{\pdag}{{\phantom{\dagger}}}

\title{Quantum criticality in the spin-${1}/{2}$ 
Heisenberg chain system copper pyrazine dinitrate}

\author
{Oliver Breunig,${}^{1}$ Markus Garst,${}^{2,3}$ Andreas Kl\"{u}mper,${}^{4}$ Jens Rohrkamp,${}^{1}$ \\ Mark M.~Turnbull,${}^{5}$ Thomas Lorenz${}^{1\ast}$\\
\\
\normalsize{${}^{1}$II. Physikalisches Institut, Universit\"at zu K\"oln, 50937 K\"oln, Germany}\\ 
\normalsize{${}^{2}$Institut f\"ur Theoretische Physik, Universit\"at zu K\"oln, 50937 K\"oln, Germany}\\
\normalsize{${}^{3}$Institut f\"ur Theoretische Physik,   Universit\"at Dresden, 01062 Dresden, Germany}\\ 
\normalsize{${}^{4}$Fachbereich C -- Physik, Bergische Universit\"at Wuppertal, 42097 Wuppertal, Germany}\\
\normalsize{${}^{5}$Carlson School of Chemistry and Department of Physics, Clark University,}\\
\normalsize{Worcester, Massachusetts 01610, USA}\\
\\
\normalsize{$^\ast$To whom correspondence should be addressed; E-mail:  tl@ph2.uni-koeln.de}
}

\date{}

\begin{document}

\baselineskip16pt


\maketitle


% The abstract should be a single paragraph, not to exceed 250 words and ideally closer to 200, written in plain language that a general reader can understand. It should include
% An opening sentence that states the question/problem addressed by the research AND
% Enough background content to give context to the study AND
% A brief statement of primary results AND
% A short concluding sentence.
% Do not include citations or undefined abbreviations in the abstract. Any abbreviations that appear in the title should be defined in the abstract.

\begin{sciabstract}
The magnetic insulator copper pyrazine dinitrate comprises antiferromagnetic spin-1/2 chains that are well described by the exactly solvable one-dimen\-sional Heisenberg model, providing a unique opportunity for a quantitative comparison between theory and experiment. Here, we investigate its thermodynamic properties with a particular focus on the field-induced quantum phase transition. Thermal expansion, magnetostriction, specific heat, magnetization and magnetocaloric measurements are found to be in excellent agreement with predictions from exact Bethe-Ansatz results as well as from effective field theory. Close to the critical field,  thermodynamics obeys the expected quantum critical scaling behavior, and, in particular, the magnetocaloric effect and the Gr\"uneisen parameters diverge in a characteristic manner. Apart from realizing a paradigm of quantum criticality, our study instructively illustrates fundamental principles of quantum critical thermodynamics.
\end{sciabstract}

\section*{Introduction}

A quantum phase transition arises when the ground state of a quantum system changes as a function of an external parameter such as pressure or magnetic field. The quantum critical fluctuations associated with this instability often give rise to exotic behaviour that is in stark contrast to the conventional properties of materials~\cite{Lohneysen2007}. They are suggested to be at the origin of the anomalous characteristics of a series of correlated electron systems like high-$T_c$ cuprates, Fe-based superconductors or heavy-fermion compounds~\cite{Sachdev2008,Gegenwart2008,Sachdev2011,Kuo2016,Mazzone2017}. These systems are so complex, however, that the underlying quantum phase transitions are often hard to identify. In this context, exactly solvable models provide an important guidance for the analysis of enigmatic quantum phase transitions in more complex systems.
 
Such models can be realized in spin systems when the interaction $J$ between localized magnetic moments is effectively restricted to one-dimensional chains, e.g.\ the spin-1/2 XXZ chain model 
\begin{equation}
\mathcal{H}= \sum\limits_{i} \left[ J \left(S_i^x S_{i+1}^x + S_i^y S_{i+1}^y + \Delta S_i^z S_{i+1}^z \right) + g\mu_B \mu_0\vec{H} \vec{S}_i \right].
\label{eq:Hamilton}
\end{equation}
Here, $\vec S_i$ is the spin-1/2 operator on site $i$, $g$ the electronic $g$-factor, and $\mu_{\rm B}$ the Bohr magneton, $\Delta$ describes an anistropy of the exchange coupling $J$ and a quantum phase transition can be induced by the magnetic field $H$. For $\Delta \gg 1$ or $\ll 1$, model~\eqref{eq:Hamilton} covers Ising- or XY-spin chains, respectively, and in both cases a transverse magnetic field, i.e. $\vec H \perp \hat z$, induces the Ising quantum phase transition, which is the most prominent textbook example of quantum criticality~\cite{Sachdev2001}. Experimental realizations are, e.g., the materials LiHoF$_4$ \cite{Bitko1996}, CoNb$_2$O$_6$ \cite{Coldea2010b} with Ising- and Cs$_2$CoCl$_4$~\cite{Kenzelmann2002a,Breunig2013} with XY-type anisotropy.

\begin{figure}[t]
  \centering
  \includegraphics[width=0.75\textwidth]{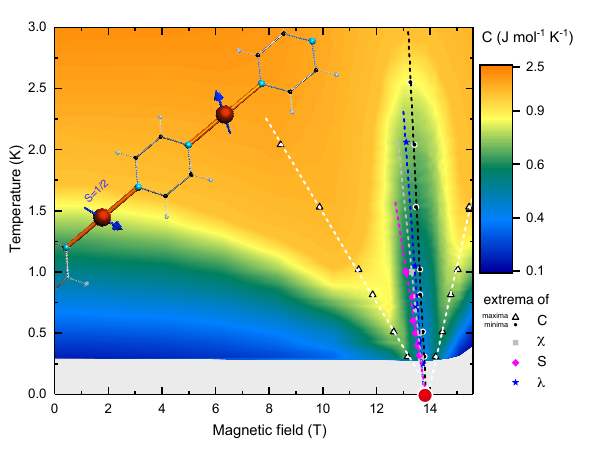}  
 \caption{{\bf Figure 1: Phase diagram of CuPzN with a field-induced quantum critical point.} The color code reflects the measured specific heat $C(H,T)$, and symbols indicate the positions of various thermodynamic signatures that obey the quantum critical scaling $T \sim |H-H_{\rm c}|^{\nu z}$ with $\nu z =1$ and $\mu_0 H_{\rm c} = 13.9$~T. Also shown is the basic structure of the Cu$^{2+}$ spin chains with $S=\frac{1}{2}$ that are exchange-coupled via pyrazine rings C$_4$H$_4$N$_2$.
}  
\label{fig1}
\end{figure}

A quantum phase transition belonging to a different universality class arises in spin-1/2 chains with isotropic Heisenberg exchange ($\Delta =1$). This model respresents one of the most fundamental strongly correlated quantum systems and the exact solution of its groundstate was pioneered by Bethe in 1931~\cite{Bethe1931}. Much later~\cite{Takahashi1971,Gaudin1971}, this was extended to finite-temperature calculations of the free energy $\mathcal{F}_{\rm 1D}(T,H)$ and then further improved in Ref.~\cite{Klumper1998}, which now allows for a precise quantitative prediction of all thermodynamic properties. Up to a critical field $g  \mu_{\rm B} \mu_0H_{\rm c}=2J$, the ground state constitutes a gapless Tomonaga-Luttinger spin liquid, while at $H\ge H_{\rm c}$ an excitation gap opens and the magnetization is fully saturated. The single-particle excitations of this saturated ground state correspond to single spin-flips, i.e., gapped magnons. On decreasing the field, these magnons condense and, due to their strong mutual interaction, form a Fermi surface for $H<H_{\rm c}$. The asymptotical thermodynamics close to quantum criticality is described by the one-dimensional free fermion dispersion $\varepsilon_k = \frac{\hbar^2 k^2}{2m} - \mu$  with mass $m = \hbar^2/(a^2 J)$, chemical potential $\mu = 2 J - g  \mu_{\rm B} \mu_0H $, and lattice constant $a$. The associated critical free energy per spin is 
\begin{align} \label{Scaling}
\mathcal{F}_{\rm cr}(T,H) = \frac{(k_B T)^{3/2}}{\sqrt{J}}  f\Big(\frac{g_b  \mu_B \mu_0(H - H_{\rm c})}{k_B T} \Big)
\end{align}
with the scaling function 
$f(x) = -\frac{\sqrt{2}}{\pi} \int_{0}^{\infty} dy \log(1 + e^{-y^2-x})$.
The linear scaling of Eq.~\eqref{Scaling}, $T \sim |H - H_{\rm c}|^{\nu z}$, implies $\nu z = 1$ with correlation-length exponent $\nu=1/2$ and dynamical exponent $z=2$.

An almost ideal material to study this quantum phase transition is copper pyrazine dinitrate (CuPzN) Cu(C$_4$H$_4$N$_2$)(NO$_3$)$_2$. It comprises spin-$1/2$ chains of Cu$^{2+}$ ions  along the $a$ axis which interact via the pyrazine rings (C$_4$H$_4$N$_2$), see inset of Fig.~\ref{fig1}, with an antiferromagnetic exchange $J/k_{\rm B} = 10.6$~K~\cite{Hammar1999,Stone2003}. Magnetic anisotropies remain negligible and result in a weak $g$ factor anisotropy~\cite{McGregor1976} with $g_b = 2.27$ for $\vec H\|b$ resulting in a critical field $\mu_0 H_{\rm c}\simeq 13.9$~T that is accessible by laboratory magnets. Interchain couplings are so weak that long-range antiferromagnetic order only develops below $T_{\rm N}= 107$~mK~\cite{Lancaster2006}. Typical signatures of quantum criticality have been reported for CuPzN~\cite{Jeong2015} based on measurements of magnetization~\cite{Kono2015}, nuclear magnetic relaxation~\cite{Kuhne2009}, thermal expansion~\cite{Rohrkamp2010} and magnetic heat transport~\cite{Sologubenko2007a}. In this report, we compare a comprehensive set of thermodynamic data of CuPzN to the analytic Bethe-Ansatz solutions of the Heisenberg chain model~\eqref{eq:Hamilton}. 

 \begin{figure*}
  \centering
  \includegraphics[width=\textwidth]{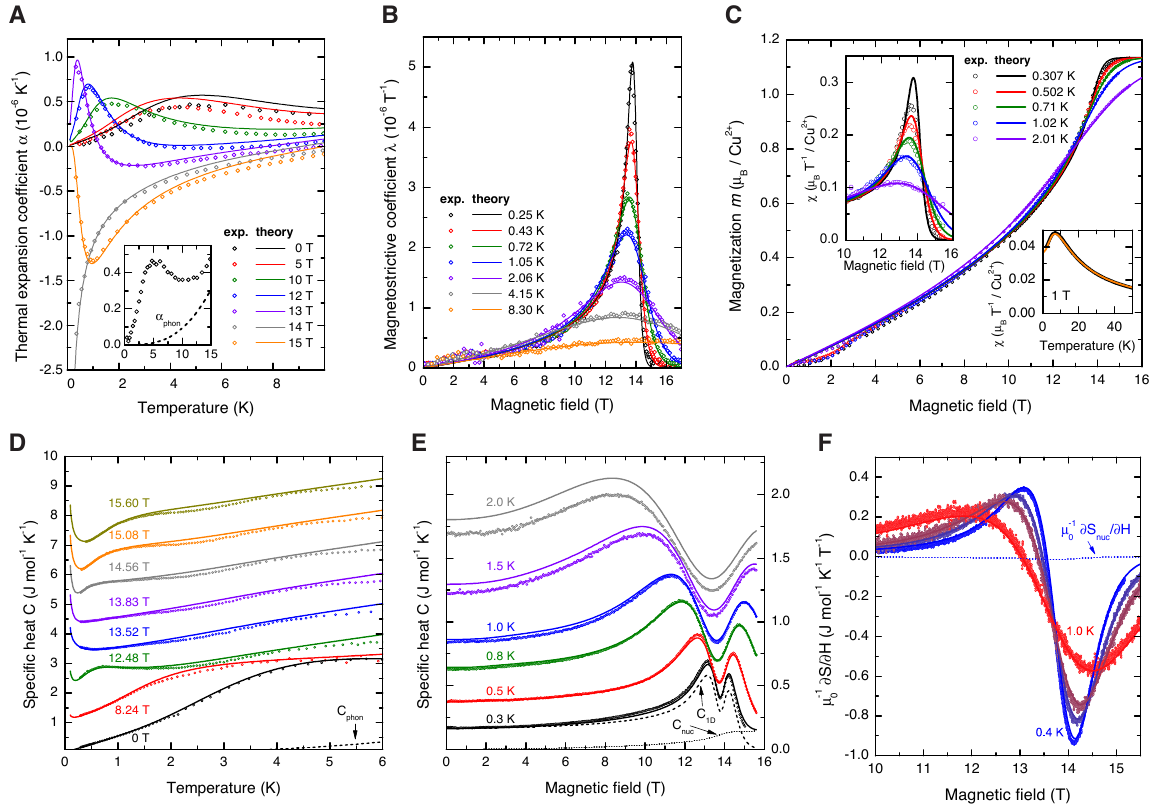}
 \caption{{\bf Figure 2: Thermodynamic quantities of the Heisenberg spin-chain compound CuPzN.} (A) thermal expansion $\alpha$, (B) magnetostriction $\lambda$, (C) magnetic moment $m$ per spin and susceptibility $\chi = \partial m/\partial (\mu_0 H)$ (insets), (D,E) molar specific heat $C$ as a function of temperature and magnetic field, respectively, and (F) the field derivative of the molar entropy $\partial S/\partial H$. In (D) and (E), the data for increasing field and temperature are offset with respect to each other by 1 and 0.1~Jmol$^{-1}$K$^{-1}$, respectively. Symbols are experimental data and solid lines are fits obtained via the Bethe-Ansatz solution of the Heisenberg spin-chain model~\eqref{eq:Hamilton} with the parameter set $J/k_{\rm B} = 10.6$~K, $\partial J/\partial p = 0.25$~K/GPa, and $g_b = 2.27$. For $\alpha$ and $C$, field-independent phononic background contributions are included, which are calculated by the Debye formula and become relevant above 5~K; see dashed lines in (A) and (D). For $C$, an additional contribution $C_{\rm nuc}$ from the nuclear spins of copper becomes relevant at lowest temperatures and high fields. For $T=0.3$~K, the calculated $C_{\rm nuc}$ is shown by the dotted line in (E), while the bare Heisenberg contribution $C_{\rm 1D}$ is displayed by the dashed line and the sum of both (solid line) reproduces the experimental data. The corresponding nuclear contribution $\partial S_{\rm nuc}/\partial H$ is negligibly small as shown by the dashed line in (F).}  
  \label{fig2}
 \end{figure*}

\section*{Results}

Figure~\ref{fig2} gives an overview of the experimental data. The $a$-axis thermal expansion $\alpha = (\partial L_a/\partial T)/L_a$ and magnetostriction $\lambda = (\partial L_a/\partial (\mu_0 H))/L_a$, partly presented already in a preliminary report~\cite{Rohrkamp2010},  are displayed as open symbols in panels (a) and (b), respectively. Below 10~K, $\alpha$ is almost entirely of magnetic origin; phonons hardly contribute anymore as is illustrated in the inset. In zero field, the magnetic contribution results in a broad maximum around 5~K which shifts to lower temperature with increasing field. A characteristic sign change of $\alpha$ is observed close to $H_{\rm c}$ reflecting entropy accumulation close to the quantum phase transition~\cite{Garst2005}. Close to the critical field, the magnetostriction $\lambda$ exhibits a strong anomaly that sharpens with decreasing temperature. The magnetic contributions of the Heisenberg chain~\eqref{eq:Hamilton} to $\alpha$ and $\lambda$ result from a pressure-dependent $J(p)$ (for uniaxial $p\|a$) and are given by $\alpha_{\rm 1D} = \frac{1}{V_S} \frac{\partial^2 \mathcal{F}_{\rm 1D}}{\partial p \partial T}$ and $\lambda_{\rm 1D} = \frac{1}{V_S}  \frac{\partial^2 \mathcal{F}_{\rm 1D}}{\partial p \partial (\mu_0 H)}$, respectively. Here, $V_S = 202$ \AA$^3$ is the volume per spin in CuPzN and $\frac{\partial \mathcal{F}_{\rm 1D}}{\partial p}=\frac{\partial \mathcal{F}_{\rm 1D}}{\partial J}\frac{\partial J}{\partial p}$.  The resulting fits based on $\mathcal{F}_{\rm 1D}(T,H)$ of the Bethe-Ansatz solution are shown by the solid lines in Fig.~\ref{fig2} (A,B). For $\alpha(T,H) =\alpha_{\rm 1D}(T,H) +\alpha_{\rm phon}(T)$, a field-independent phononic background based on the Debye model (see Inset) has been included. Both, $\alpha(T,H)$ and $\lambda(T,H)$ are well reproduced, apart from some minor deviations of the low-field $\alpha(T)$ around 5~K, which may partly arise from an improper description of $\alpha_{\rm phon}(T)$.  Remarkably, the quantum critical signatures around $H_{\rm c}$ are perfectly reproduced, although there is essentially only one adjustable parameter $\partial J/\partial p = 0.25$~K/GPa, because $J/k_{\rm B} = 10.6$~K and $g_b=2.27$ are known from previous studies~\cite{Hammar1999,McGregor1976}. Note that this pressure dependence is more than one order of magnitude smaller than the corresponding values of the spin-Peierls system CuGeO$_3$~\cite{Buchner1996}. This suggests that the magnetoelastic coupling in CuPzN is small enough that the magnetic order at 107~mK preempts a spin-Peierls transition, which is an inherent instability of half-integer spin chains towards a combined lattice and spin dimerization~\cite{Bray1983,Inagaki1983}.

The magnetization of CuPzN is compared to the Bethe-Ansatz solutions in Fig.~\ref{fig2}(C). At 0.3~K the magnetic moment per spin $m$ has a relatively sharp kink close to $\mu_0 H_{\rm c}\simeq 13.9$~T and reaches saturation above about 15~T, which causes an asymmetric peak in the differential susceptibility $\chi = \partial m/\partial (\mu_0 H)$ shown in the upper inset. With increasing temperature, the critical signatures systematically broaden and the data are well described by the Heisenberg model (solid lines), although the agreement at lowest temperature is not as good as that of $\lambda$. The lower inset shows that the model also reproduces $\chi(T, \mu_0 H = 1\;{\rm T})$ up to high temperature.

The molar specific heat $C$ as a function of temperature and field is displayed in Fig.~\ref{fig2}(D) and (E), respectively. At zero field, $C(T)$ strongly resembles the thermal expansion. This is rooted in the single energy scale $J$ of the Heisenberg model~\eqref{eq:Hamilton} that implies a Gr\"uneisen scaling $(\alpha/C)|_{H=0} = \frac{1}{V_m} \frac{\partial \ln J}{\partial p}$ with the molar volume $V_m = N_{\rm A} V_S$. The low-temperature $C(H)$ is characterized by a slightly asymmetric double-peak structure centered at $H_{\rm c}$ that broadens with increasing temperatures. Such a double peak is, in fact, generic for metamagnetic quantum criticality~\cite{Weickert2010}. The positive curvature $\partial^2C/\partial H^2$ at $H_{\rm c}$ is linked via a Maxwell relation to the curvature of the %differential 
susceptibility $\partial^2 \chi/\partial T^2$ that is positive because of the diverging $\chi(T\rightarrow 0, H=H_{\rm c})$. 

The behavior of $C(T,H)$ is dominated by the magnetic contribution of the Heisenberg chains, but for its quantitative description we have to consider also contributions from phonons and from nuclear spins. While the phonons start to contribute above about 5~K, the nuclear contribution is relevant at lowest temperatures and high fields only, as is shown exemplarily for $T=0.3$~K by the dotted line in Fig.~\ref{fig2}(E). The calculated total specific heat is then given by the solid lines, which perfectly reproduce the experimental data up to 1~K, while some systematic deviations on the order of 10\% are found around 2.5~K, whose origin remains unclear (see supplementary material).

Finally, in Fig.~\ref{fig2}(F) we present the isothermal magnetocaloric effect, i.e., the field derivative of the molar entropy $\partial S/\partial H$. Similar to $\alpha$, this quantity shows a characteristic sign change approaching a divergence on decreasing temperature, which directly reflects the entropy accumulation close to $H_{\rm c}$. Again the experimental data are fully reproduced by the Bethe-Ansatz solution of~\eqref{eq:Hamilton}; solid lines in Fig.~\ref{fig2}(F). Note that there is also a contribution from the nuclear spin entropy, shown by the dotted line, but it is so small that it can be safely neglected.

\section*{Discussion}

 \begin{figure*}
  \centering
  \includegraphics[width=\columnwidth]{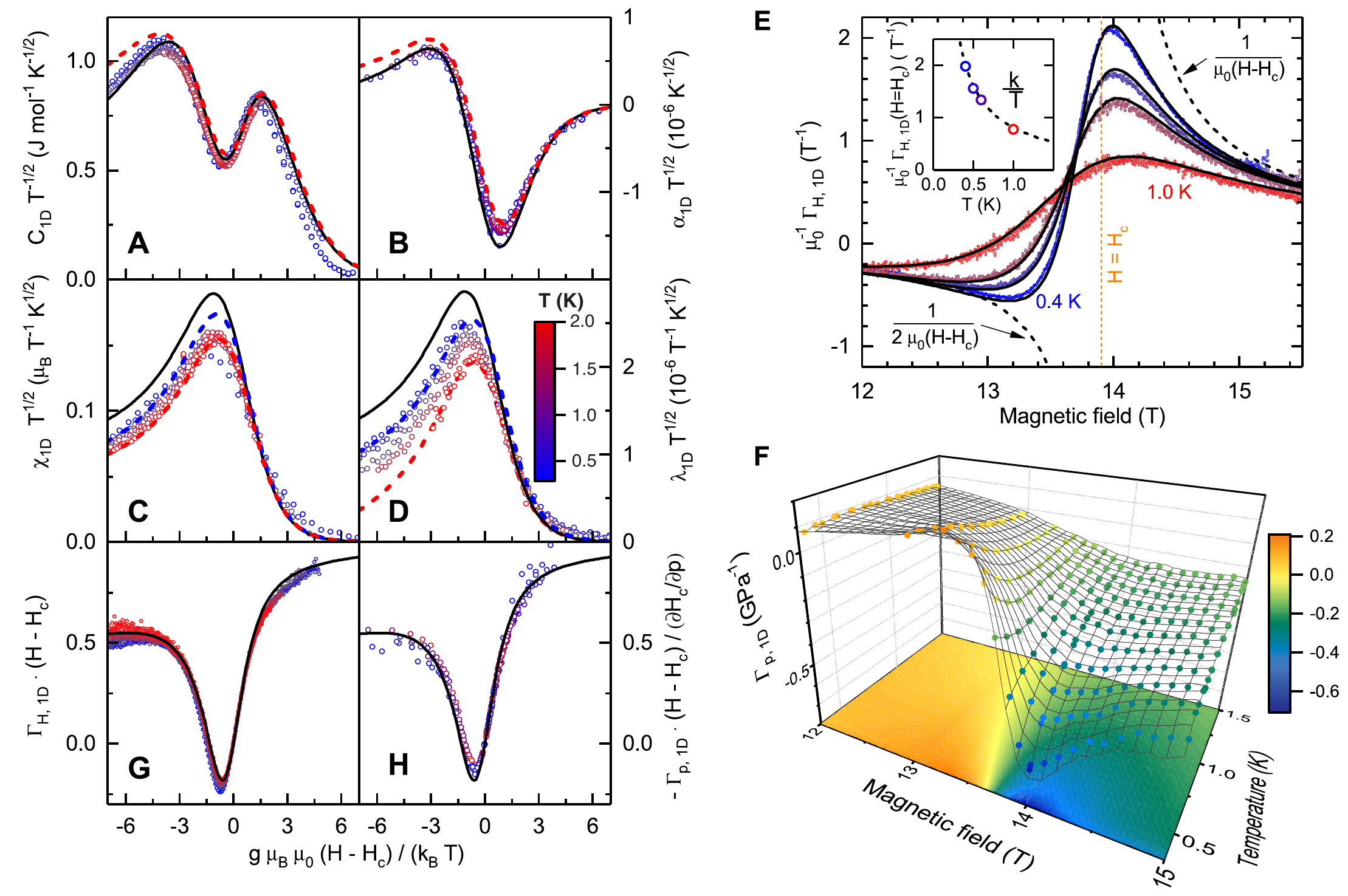}
 \caption{{\bf Figure 3: Quantum critical scaling of thermodynamic quantities close to the critical field $\mathbf{H_{\rm c}}$.} Non-critical background contributions due to phonons and/or nuclear spins have been subtracted. Multiplication by $\sqrt{T}$ and plotting versus the scaling parameter $g_b \mu_B \mu_0 (H-H_{\rm c})/(k_B T)$ causes a collapse of (A) $C_{\rm 1D}/T$, (B) $\alpha_{\rm 1D}$, (C) $\chi_{\rm 1D}$, and (D) $\lambda_{\rm 1D}$ towards critical scaling functions (solid black lines) which are derived from $f(x)$ of Eq.~\eqref{Scaling}; symbol colors indicate different temperatures from 0.3\,K (blue) to 2.0\,K (red). The corresponding Bethe-Ansatz results calculated for $T=2$\,K are shown as red dashed lines in (A-D) and, additionally, for $T=0.25$\,K as blue dashed lines in (C,D). The importance of corrections to scaling increases from panels (A) to (D) spoiling a complete scaling collapse. (E) The experimentally obtained magnetic-field dependent Gr\"uneisen parameter $\Gamma_{\rm H, 1D}$ (symbols) is perfectly described by its critical behavior (solid lines) given by Eq.~\eqref{eq:Gamma_scal}. The dashed lines show the universal divergences of Eq.~\eqref{eq:Gamma_scal} in the zero-temperature limit and the inset compares the corresponding ${\rm k}/T$ divergence at $H=H_{\rm c}$ with the experimental data (symbols).  (F) The pressure-dependent Gr\"uneisen parameter $\Gamma_{\rm p, 1D}$ is, according to Eq.~\eqref{eq:Gamma_H_p}, proportional to $\Gamma_{\rm H, 1D}$  and consequently both Gr\"uneisen parameters collapse on the very same scaling function $\Phi(x)$ of Eq.~\eqref{eq:Gamma_scal} as shown in (G,H).}
  \label{fig3}
 \end{figure*}

We now turn to a discussion of the field-induced quantum criticality and compare the data with the scaling predictions of Eq.~\eqref{Scaling}. For this, we confine ourselves to data obtained below 2~K in the field range $\mu_0 H_{\rm c} \pm 4$ T. In order to extract the bare magnetic properties of the Heisenberg spin chains, phononic and/or nuclear background contributions are subtracted. From the full fits of Fig.~\ref{fig2} it is, however, inferred that $C_{\rm nuc}$ causes the only relevant correction in this low-temperature range. According to Eq.~\eqref{Scaling}, susceptibility $\chi_{\rm 1D}$, specific heat coefficient $C_{\rm 1D}/T$, thermal expansion  $\alpha_{\rm 1D}$, and magnetostriction $\lambda_{\rm 1D}$ are all predicted to diverge as $1/\sqrt{T}$ at the critical field. After multiplying by $\sqrt{T}$, these quantities are described  by universal scaling functions asymptotically close to the quantum critical point when plotted versus the scaling variable $g_b \mu_B \mu_0 (H-H_{\rm c})/(k_B T)$. These scaling functions are directly related to $f$ of Eq.~\eqref{Scaling} and are shown as solid black lines in Fig.~\ref{fig3}(A)-(D). We find a very good scaling collapse for $C_{\rm 1D}$ and $\alpha_{\rm 1D}$, but substantial deviations are observed for $\chi_{\rm 1D}$ and are even more pronounced for $\lambda_{\rm 1D}$. 
These deviations arise from corrections to scaling which, depending on their relative magnitude, can spoil a full scaling collapse in an extended parameter range. This is confirmed by the blue and red dashed lines that display the Bethe-Ansatz solutions for temperatures $0.25$~K and $2$~K, respectively. As discussed in the supplementary material, these corrections to scaling are only negligible in the limit $\sqrt{k_B T/J} \ll 1$ whereas $\sqrt{k_B T/J} = 15\%$ is still sizeable even at $T = 0.25$~K. The good scaling collapse observed for $C_{\rm 1D}$ and $\alpha_{\rm 1D}$ is attributed to numerical factors that are small although formally of order one.

A quantity of particular interest close to field-induced quantum criticality is the adiabatic magnetocaloric effect defined by the magnetic-field dependent Gr\"uneisen parameter $\Gamma_{\rm H} = - (\partial S_{\rm H}/\partial H)/C_{\rm H} = \frac{1}{T} (\partial T/\partial H)|_S$  that quantifies the ability of the system to adiabatically change the temperature upon a field change.
General scaling considerations predict that $\Gamma_{\rm H}$ diverges with characteristic exponents close to quantum criticality, which allows one to identify and classify the quantum critical point~\cite{Zhu2003,Garst2005}. Fig.~\ref{fig3}(E) shows that with decreasing temperature  the obtained $\Gamma_{\rm H, 1D}(T,H)$ of CuPzN approaches a sign-change singularity at $H_{\rm c}$, in agreement with the expected asymptotic quantum critical behavior
\begin{align}
\Gamma^{\rm c}_{\rm H, 1D} = \frac{1}{H-H_{\rm c}} \Phi\Big(\frac{g_b  \mu_B \mu_0 (H-H_{\rm c})}{k_B T}\Big)\; .
\label{eq:Gamma_scal}
\end{align}
The scaling function is related to $f$ of Eq.~\eqref{Scaling} via $\Phi(x) = 2 x (-f'(x) + 2 x f''(x) ) /[3f(x) - 4 x (f'(x) - x f''(x))]$. The critical $\Gamma^{\rm c}_{\rm H, 1D}(T,H)$ is plotted as solid lines in Fig.~\ref{fig3}(E) and perfectly reproduces the experimental $\Gamma_{\rm H, 1D}(T,H)$. The asymptotics for $x\to \pm\infty$ result in characteristic zero-temperature divergencies $\Gamma^{\rm c}_{\rm H, 1D}\sim (H-H_{\rm c})^{-1}$ with the universal prefactors $\Phi(x\to \pm\infty)=1$ and $1/2$~\cite{Zhu2003}, respectively, as shown by the dashed lines.  The data at 0.4~K are already close to this universal behavior, but only for $H< H_{\rm c}$. Close to the critical field, $\Phi(x) \sim \mathcal{C} x$ with $\mathcal{C} \simeq 0.527$, that results in the  divergence $\mu_0^{-1}\Gamma_{\rm H, 1D}(T,H_{\rm c})={\rm k}/T$ with ${\rm k} = \mathcal{C}g_b  \mu_B/(k_{\rm B} T)\simeq 0.804$~K/T, which perfectly agrees with the data; see inset of Fig.~\ref{fig3}(E). 

Closely related to $\Gamma_{\rm H}$ is the pressure-dependent Gr\"uneisen parameter $\Gamma_{\rm p} = V_m \alpha/C$ \cite{Zhu2003,Garst2005}. The most singular contribution to $\alpha_{\rm 1D}$ arises from the pressure dependent $H_{\rm c}(p)$ so that asymptotically $\alpha_{\rm 1D} V_m= - \partial S_{\rm 1D}/\partial p = (\partial S_{\rm 1D}/\partial H) (\partial H_{\rm c}/\partial p)$. This yields the proportionality
\begin{align}
\Gamma^{\rm c}_{\rm p,1D} = - \frac{\partial H_{\rm c}}{\partial p} \Gamma^{\rm c}_{\rm H, 1D} 
= - \frac{2}{g_b \mu_{\rm B} \mu_0} \frac{\partial J}{\partial p} \Gamma^{\rm c}_{\rm H, 1D} 
\label{eq:Gamma_H_p}
\end{align}
close to quantum criticality. The experimentally obtained $\Gamma_{\rm p,1D} = V_m \alpha_{\rm 1D}/C_{\rm 1D}$ is displayed in Fig.~\ref{fig3}(F) and already indicates that apart from the opposite signs the field and temperature dependences of $\Gamma_{\rm p,1D}$ and $\Gamma_{\rm H,1D}$ are identical. This is quantitatively confirmed in Fig.~\ref{fig3}(G,H) showing that indeed both, $\Gamma_{\rm H,1D}\,(H-H_{\rm c})$ as well as $-\Gamma_{\rm p,1D}\,(H-H_{\rm c})\,/\,(\partial H_{\rm c}/\partial p)$ perfectly collapse on the very same scaling function $\Phi(x)$ from Eq.~\eqref{eq:Gamma_scal}.

In summary, the low-temperature thermodynamics of CuPzN is excellently described by the Heisenberg spin-1/2 chain model after taking into account small phononic and/or nuclear background contributions. We have demonstrated the emergence of universal scaling behavior close to its field-induced quantum critical point. Comparison between experiment and exact Bethe-Ansatz solution has elucidated the importance of corrections to scaling, which varies from quantity to quantity and might spoil a scaling collapse over an extended parameter regime. 
%Our study emphasizes the significance of such scaling corrections for the reliable identification of quantum critical exponents from experimental data.
%
Our study establishes CuPzN as a paradigm of quantum criticality that can serve as a reference in the quest for understanding putative quantum critical behavior in other strongly correlated systems.

\section*{Materials and Methods}

\subsection*{Sample Preparation and Measurements}
Single crystals  of CuPzN were grown from an aqueous solution of pyrazine and Cu nitrate via slow evaporation. Typical crystals have a length along the $a$ axis of about 10 mm. Perpendicular to the $a$ axis the crystals are usually smaller than 1 mm with $b$ being the shortest axis. Crystals of CuPzN are orthorhombic ($Pmna$) with the lattice constants $a=6.712$ \AA, $b=5.142$~\AA\ and $c=11.73$~\AA\ \cite{Santoro1970}. Magnetic fields were applied along the crystallographic $b$ axis. Measurements of the thermal expansion and the magnetostriction were performed using a home-built capacitance dilatometer in a transverse configuration, i.e., measuring the length change along the chain direction $a$ with the magnetic field applied along $b$. The uniaxial thermal expansion coefficient $\alpha$ and the magnetostrictive coefficient $\lambda$ of the $a$ axis were obtained from the data by numerical differentiation, $(\alpha,\lambda)=\frac{1}{L_a} \frac{\partial\Delta L_a}{\partial (T,\mu_0 H)}$. The specific heat was measured using a home-built calorimeter based on the relaxation time method. The addenda was obtained in a separate run and subtracted from the obtained total specific heat. The magnetization was measured with a capacitive Faraday magnetometer that was previously calibrated in magnetic fields and matched to the data taken at temperatures larger than 2 K with the VSM option of a commercial PPMS system (Quantum Design). The magnetocaloric effect was measured in a continuous way as described in the supplementary material.

\subsection*{Theoretical Modeling}
For the calculation of the thermodynamical potential of the spin-1/2 Heisenberg chain we use the method described in Ref.~\cite{Klumper1998}. This requires the numerical solution of a set of just two non-linear integral equations (NLIEs) for two auxiliary functions. There are equivalent but numerically differently conditioned formulations of these NLIEs. Here we use the formulation of Ref.~\cite{Kundu2003}.

The free energy per site for temperature $T$ and magnetic field $H$ is obtained as a contour integral
\begin{equation}
\mathcal{F}_{\rm 1D}(T,H)=\frac{J-2g \mu_B \mu_0 H}{4}
-\frac {k_BT}{2\pi}\int_{{\cal C}} \frac{\log(1+a(y))}{(y+{\rm i})y}dy\, .    \label{eigenvalue}
\end{equation}
Here, ${\cal C}$ is a narrow closed contour around the entire real axis involving an auxiliary function $a(x)$. This function satisfies the NLIE
\begin{equation}
\log a(x)= \frac{J}{2k_BTx(x+{\rm i})}-\frac{g \mu_B \mu_0 H}{k_BT} -\frac 1{\pi}\int_{{\cal C}} 
\frac{\log(1+a(y))}{(x-y)^2+1}dy\,. \label{NLIEa}
\end{equation}
In this formulation, the invariance of the free energy under a sign change of the magnetic field ($H\to-H$) is not manifest, but of course true. The NLIE (\ref{NLIEa}) can be solved iteratively with fast convergence for positive values of $H$. In numerical calculations the integral over a function $g(x)$ along the contour ${\cal C}$ is replaced by integrals over two functions $g(x+{\rm i}/2)$ and $g(x-{\rm i}/2)$ along the real axis. In this manner, the single contour NLIE is equivalent to two coupled NLIEs. Convolutions are treated by Fast Fourier algorithms.

\section*{Supplementary Material}
section A. Magnetocaloric effect \\
section B. Deviations of $C$ around 2.5~K \\
section C. Nuclear contributions \\
section D. Quantum critical theory and corrections to scaling \\
Fig. S1: Magnetocaloric effect measurement. \\
Fig. S2: Theoretical prediction for the scaling of quantum critical thermodynamics in CuPzN. \\
Fig. S3: Deviations from critical scaling. \\
references \cite{Rost2009,Tokiwa2011,Wolf2011,Dubiel2009}

%\bibliography{CuPzN_170221_library.bib}
%\bibliographystyle{ScienceAdvances}

\noindent \textbf{Acknowledgements:} 

\noindent \textbf{Funding:} This work was supported by the Deutsche Forschungsgemeinschaft via FOR 960 (Quantum Phase Transitions), SFB 1143 (Correlated Magnetism: From Frustration To Topology), and CRC 1238 (Control and Dynamics of Quantum Materials; Project No. B01). O.B. acknowledges support from the Quantum Matter and Materials Program at the University of Cologne funded by the German Excellence Initiative.\\
\noindent \textbf{Author Contributions} T.L. initiated the investigations on CuPzN and coordinated the project. M.M.T. provided the single crystals, O.B. and J.R. performed the measurements. M.G. headed the analysis of the data. A.K. contributed the numerical solutions of the Bethe-Ansatz model, O.B. further processed these solutions in order to adjust them to the experimental data.  O.B., M.G., and T.L. wrote the manuscript with input from all authors.
\\
\noindent \textbf{Competing Interests} The authors declare that they have no competing financial interests.\\
\noindent \textbf{Data and materials availability:} All data needed to evaluate the conclusions in the paper are present in the paper and/or the Supplementary Materials. Additional data related to this paper may be requested from the authors.

\newpage

\section*{Supplementary Materials for: ''Quantum criticality in the \\ spin-$1/2$ Heisenberg chain system copper pyrazine dinitrate``}

\section*{section A. Magnetocaloric effect}

By definition, the magnetic Gr\"uneisen parameter 
	$\Gamma_H=-\frac{\left.\frac{\partial M}{\partial T}\right|_H}{C_H}$
	can be determined from the ratio of the temperature-derivative of the magnetization $M$ and the specific heat $C_H$, both measured at fixed values of the magnetic field $H$. Alternative experimental methods to determine $\Gamma_H$ are described, e.g., in Refs.~\cite{Rost2009,Tokiwa2011,Wolf2011}. Here, we chose to measure the isothermal entropy change $\left.\frac{\partial S}{\partial H}\right|_T=\left.\frac{\partial M}{\partial T}\right|_H$ in a continuous way while sweeping the magnetic field.  The experimental setup consists of a standard thermal relaxation-time calorimeter where the sample is fixed to the sample platform using a small amount of Apiezon N grease. While sweeping the magnetic field, the sample temperature $T$ is kept at a constant difference $\Delta T$ above the bath temperature $T_0$ by adjusting the power $P$ applied to the sample heater at the platform, which is coupled by the thermal conductance $K$ to the heat bath (cf.~Fig.~{S1}). 
Under isothermal conditions, the entropy change $dS$ is given by its field dependence 	 
   \begin{equation}\label{eqn:theory-mce-dE}
	 T\,dS = T\left.\frac{\partial S}{\partial H}\right|_T dH = (P-K\,\Delta T)\,dt\, ,
	\end{equation}
	which is balanced by a variation of the heating power $P$. For reversible processes 
	\begin{equation}
	  P_\gamma(H)=K\,\Delta T + \gamma\, T\left.\frac{\partial S}{\partial H}\right|_T\,,
	\end{equation}
where $\gamma=\frac{dH}{dt}$ denotes the magnetic-field sweep rate. As expected $P_\gamma(H)$ is antisymmetric with respect to the field-sweep direction, see Fig.~{S1}, and the isothermal magnetic-field dependence of the entropy is given by
	\begin{equation}
	  \left.\frac{\partial S}{\partial H}\right|_T = \frac{P_\gamma(H)-P_{-\gamma}(H)}{2\left|\gamma\right| T}.
	\end{equation}
	The magnetic-field dependent Gr\"uneisen parameter $\Gamma_H$ is then obtained by additionally dividing by the heat capacity as a function of $H$, which was measured in a separate run in the same setup using the thermal relaxation time method. 
	
\begin{figure}[t]
  \centering
  \includegraphics[width=.6\columnwidth]{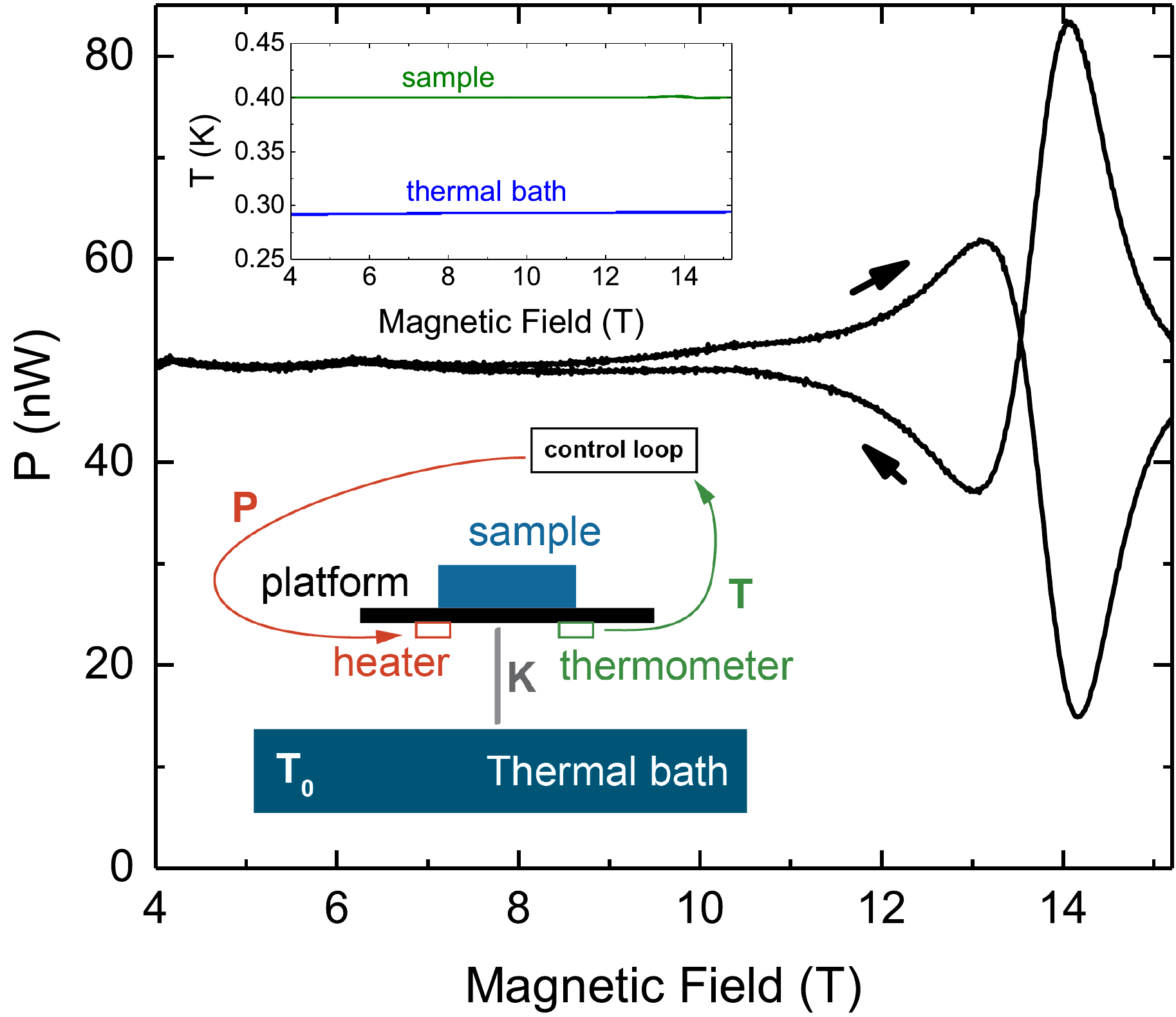}
 \caption{{\bf Fig. S1: Magnetocaloric effect measurement.} Raw data of the heating power P applied to the sample heater during a magnetic field sweep with increasing and decreasing magnetic field (indicated by arrows) at a rate of $\pm 0.2\,$T/min, acquired with a setup as schematically shown in the main panel. While sweeping the field, the temperatures of the sample and of the thermal bath (shown in the inset) are constant within 1\% at 0.29 and 0.4\,K, respectively, which confirms the precise tuning of the temperature control loop.}
  \label{figmke}
\end{figure}

\section*{section B. Deviations of $\mathbf{C}$ around 2.5~K}

Around 2.5~K, the specific heat calculated for the Heisenberg model deviates by up to 10\,\% from the experimental data. These deviations cannot be explained by phonon or nuclear contributions because the bare magnetic contribution $C_\mathrm{1D}$ is already larger than the total measured specific heat. This experimental result was reproduced by measurements on different CuPzN crystals from different growth procedures. Moreover, it has been independently obtained with our home-built low-temperature calorimeter and by using the specific heat option of the commercial PPMS system (Quantum Design). In order to fit the experimental data in this temperature range with the 1D Heisenberg model one would have to increase the exchange coupling $J$ by 10\%, but this disagrees with the fits of all other (thermodynamic) measurements and would also decrease the overall agreement of the general $C(T,H)$ data. 

\section*{section C. Nuclear contributions}
Nuclear contributions can be identified in the specific heat data at low temperature $T\lesssim 0.5$~K. They arise from the nuclear spins ($I=3/2$) of the copper atoms in the compound due to a finite splitting of energy levels by hyperfine interactions and their thermal occupation. Here, we restrict ourselves to interactions with the external magnetic field $\mu_0H$ and with the hyperfine field arising from the spins of the surrounding unpaired electrons, which is proportional to the magnetic moment $m_\mathrm{1D}$ per spin of the Heisenberg chain. The nuclear free energy per atom is 
\begin{equation} 
\mathcal{F}_\mathrm{nuc}= -k_{\rm B}T \log\left[\sum\limits_{I=-3/2}^{I=3/2}e^\frac{-g_N \mu_\mathrm{N}B_\mathrm{hf} I}{k_\mathrm{B} T}\right],
\label{fnuc}
\end{equation}
with the natural abundance-averaged nuclear g factor $g_N\simeq 1.516$ of copper, the nuclear magnetic moment $\mu_\mathrm{N}$, and the effective total hyperfine field $B_\mathrm{hf}=\mu_0H+A m_\mathrm{1D}$. Note that the single adjustable parameter of Eq.~(\ref{fnuc}) is $A$, which describes the coupling of the net magnetization to the hyperfine field.
Within the temperature range of the present experiment
$T\gg g_N \mu_N B_\mathrm{hf} / k_\mathrm{B}$ ($\approx 0.05\,\mathrm{K}$ for a typical value $B_\mathrm{hf}\approx 10^2\,\mathrm{T}$) only the high-temperature tail of a Schottky anomaly contributes to the molar specific heat,
\begin{equation}
C_\mathrm{nuc}/(N_{\rm A}k_\mathrm{B}) \approx B_\mathrm{hf}^2\, g_N^2\, \mu_N^2  (I + I^2)/(3\, k_\mathrm{B}^2\, T^2).
\label{eq:cnuc}
\end{equation}
Fitting our low-temperature data with the sum $C=C_\mathrm{1D}+C_\mathrm{nuc}$ of the Heisenberg spin chain (see main text) and the nuclear contribution we obtain  $A= 46\, \mathrm{T}\mu_\mathrm{B}^{-1}$, which is comparable to the values found for other transition metal systems \cite{Dubiel2009}. Apparently, the hyperfine field $B_\mathrm{hf}$ is mainly determined by the spin chain's magnetization, which is reflected in the nuclear heat capacity's strong resemblance to the magnetization (cf.~Fig.~2(C,E)). 

A sizable nuclear contribution is only found in the specific heat at 0.3 and 0.5\,K. Nuclear contributions are neither seen in $\alpha $ and $\lambda$ nor in $m$ at any temperature. For $\alpha $ and $\lambda$, this results from neglibly small pressure dependencies of $\mathcal{F}_\mathrm{nuc}$ and for $m$ it is due to the smallness of $\mu_N \ll \mu_\mathrm{B}$. In case of the magnetocaloric effect the critical contribution
\begin{equation}
 \Gamma_{H,1D}=\frac{\partial S/\partial H - \partial S_\mathrm{nuc}/\partial H}{C-C_\mathrm{nuc}-C_\mathrm{ph}}
\end{equation}
actually differs from the directly measured total $\Gamma_H=\frac{\partial S/\partial H}{C}$. The only relevant difference, however,  arises from the same nuclear contribution $C_\mathrm{nuc}$ to the heat capacity discussed above. In the relevant low-temperature range the phonon contribution $C_\mathrm{ph}$ can be safely neglected and also the nuclear contribution to the entropy change $\partial S_\mathrm{nuc}/\partial H$ turns out to be negligibly small (see dotted line in Fig.~2(F) of the main text).

\section*{section D. Quantum critical theory and corrections to scaling}

In this section we review the quantum critical theory describing the field-induced quantum phase transition of the Heisenberg chain. Moreover, we discuss the leading  corrections to scaling for the various thermodynamic quantities measured in our experiment. In particular, we address the  observation that the deviations from the critical scaling curve is larger for the susceptibility and the magnetostriction as compared to the specific heat and thermal expansion, see Fig.~3 of the main text. We show that this is a matter of numerical coefficients that are formally of order one but turn out to be larger for the former than for the latter two quantities. 

\subsection*{D1. Free energy per spin close to quantum criticality}

The Bethe-Ansatz approach yields the free energy of the Heisenberg spin chain in terms of the non-linear integral equations~(5) and (6) in the main manuscript. An asymptotic analysis of these equations provides the following expression for the free energy per spin close to quantum criticality
\begin{align}\label{FAsymptotics}
\mathcal{F} = \frac{J - 2 g \mu_{\rm B} \mu_0 H}{4} + \frac{(k_{\rm B} T)^{3/2}}{\sqrt{J}} f_0\Big(\frac{g\mu_{\rm B} \mu_0 (H-H_c)}{k_{\rm B} T}\Big) + \frac{(k_{\rm B} T)^2}{J} f_1\Big(\frac{g\mu_{\rm B} \mu_0 (H-H_c)}{k_{\rm B} T}\Big) + \dots
\end{align}
where the functions $f_0$ and $f_1$ are given by 
\begin{align}
f_0(x) &= - \frac{\sqrt{2}}{\pi} \int_0^\infty dy \log(1 + e^{-y^2 - x}),\qquad
f_1(x) = - \frac{1}{2} f_0(x) f_0'(x)
\end{align}
with the derivative $f_0'(x) = \frac{d f_0(x)}{dx}$.

In the following, we explain that the leading terms of Eq.~\eqref{FAsymptotics} have an intuitive interpretation and can be derived by elementary methods.  Performing a standard Jordan-Wigner transformation for the spins, the Heisenberg Hamiltonian of Eq.~(1) in the main text at $\Delta = 1$ can be written in the form
\begin{align} \label{Ham}
%\mathcal{H} = \sum_i \Big( - \frac{J}{2} (c^\dagger_i c^\pdag_{i+1} + c^\dagger_{i+1} c^\pdag_{i}) + g \mu_{\rm B} \mu_0 H (c^\dagger_{i} c^\pdag_{i} - \frac{1}{2}) + J   (c^\dagger_{i+1} c^\pdag_{i+1} - \frac{1}{2}) (c^\dagger_{i} c^\pdag_{i} - \frac{1}{2}) \Big)
\mathcal{H} = \sum_i \left( - \frac{J}{2} \left(c^\dagger_i c^\pdag_{i+1} + c^\dagger_{i+1} c^\pdag_{i}\right) + J   \left(n_{i+1} - \frac{1}{2}\right) \left(n_{i} - \frac{1}{2}\right) + g \mu_{\rm B} \mu_0 H \left(n_{i} - \frac{1}{2}\right) \right)
\end{align}
where $n_i = c^\dagger_i c^\pdag_{i}$. 
In the field-polarized state at large $H$, the density of spinons described by the fermionic annihilation operator $c_i$ on site $i$ is dilute. The first term in Eq.~\eqref{FAsymptotics} just corresponds to the energy of the field polarized ground state that is empty of spinons. The second term in Eq.~\eqref{FAsymptotics} is recovered by the thermal excitation of non-interacting spinons. At low temperatures, spinons with energy $\varepsilon_k = - J \cos (a k) - (J - g\mu_{\rm B} \mu_0 H)$, where $a$ is the lattice constant, only contribute here  for small wavevectors $\varepsilon_k \approx \frac{J}{2} (a k)^2 - (2J - g\mu_{\rm B} \mu_0 H)$, which identifies the mass $m = \hbar^2/(J a^2)$ and the chemical potential $\mu = g\mu_{\rm B} \mu_0 H_c - g\mu_{\rm B} \mu_0 H$ with the critical field $g\mu_{\rm B} \mu_0 H_c = 2 J$. The integral in the $f_0$ function arises from the summation over momentum states after substituting $y = \lambda_T k$ with the thermal wavelength $\lambda_T = \frac{\hbar}{\sqrt{2 m k_{\rm B} T}}$. This second term in Eq.~\eqref{FAsymptotics} governs the low-energy asymptotics close to criticality, as discussed in the context of Eq.~(2) in the main text. 

Finally, the third term in Eq.~\eqref{FAsymptotics} defines the leading correction to scaling and derives from the interaction $J n_{i+1} n_i$ between spinons in Eq.~\eqref{Ham}. Up to a factor of $1/2$, it is already obtained by treating this interaction in first-order perturbation theory and taking the low-temperature limit. The product $f_0 f'_0$ in the definition of the $f_1$ function can be identified with a product of two momentum integrals whose integrands contain Fermi functions quantifying the occupation probability of spinons. For the $f_0$ function this becomes apparent after an integration by parts,
\begin{align}
f_0(x) = - \frac{\sqrt{2}}{\pi} \int_0^\infty dy \frac{2 y^2}{1 + e^{y^2 + x}}.
\end{align}
The additional factor $y^2$ in the integrand arises from the Pauli principle and reflects that two spinons cannot simultaneously occupy the same state. 
In order to obtain the correct numerical prefactor for the leading correction to scaling, however, one has to take into account the full two-spinon $T$-matrix, i.e., the whole series of ladder diagrams in the two-spinon sector must be summed up. The resulting $T$-matrix is given by $T(E) = J/(1 - J \Pi(E))$ with 
\begin{align} \label{Pi}
\Pi(E) = a \int \limits^{\pi/a}_{-\pi/a} \frac{dk'}{2\pi} \frac{2 \sin^2(k' a)}{E - 2 \varepsilon_{k'}}.
\end{align}
In the low-temperature limit, we can approximate the on-shell energy $E$ by the energy of two spinons at zero wavevector, i.e., $E = \varepsilon_k + \varepsilon_{-k} \approx 2 \varepsilon_0 = -2(2J - g\mu_{\rm B} \mu_0 H)$. The integral in Eq.~\eqref{Pi} then simplifies to $\Pi = - 1/J$ and the on-shell $T$-matrix becomes $T = J/2$ yielding the additional factor of $1/2$. The repeated scattering between two spinons thus reduces by half the correction to scaling.

\subsection*{D2. Critical thermodynamics}

The theoretical prediction for the quantum critical thermodynamics of CuPzN is shown for temperatures $T  = 2$~K and $T = 0.25$~K in Fig.~{S2} that compares the full Bethe-Ansatz result (dashed lines) with the critical scaling curve (black solid line) deriving only from the second term in Eq.~\eqref{FAsymptotics} and with the curves that include the leading correction to scaling deriving from the third term in Eq.~\eqref{FAsymptotics} (red and blue solid lines).

The leading correction to scaling, i.e., the third term in Eq.~\eqref{FAsymptotics} is systematically smaller than the second term by a factor of $\sqrt{k_{\rm B} T/J}$. Scaling close to criticality is only expected in the limit $\sqrt{k_{\rm B} T/J} \ll 1$ when the correction to scaling is negligible. For CuPzN $J/k_{\rm B} = 10.6$~K so that $\sqrt{k_{\rm B} T/J} \approx 43\%$ at a temperature $T = 2$~K and still $\sqrt{k_{\rm B} T/J} \approx 15\%$ at $T = 0.25$~K, which is the lowest temperature where we have performed measurements. As a result, the scaling corrections are in general expected to be sizeable in our experimental data. 
This explains the deviations from the universal scaling curve observed for the susceptibility and the magnetostriction in panel (C) and (D), respectively, of Fig.~{S2} as well as Fig.~3 of the main text. As we will show in the following, small numerical coefficients further suppress the leading correction to scaling in the other thermodynamic quantities explaining the fortuitously good scaling collapse of the specific heat, thermal expansion, magnetocaloric effect and Gr\"uneisen parameter, see Fig.~3 of the main text.

\subsubsection*{D2a. Specific heat}

The molar specific heat is defined as $C = - N_{\rm A} T\partial^2_T \mathcal{F}$ with the Avogadro constant $N_{\rm A}$.
The critical scaling part and the leading scaling correction can be cast in the scaling form
\begin{align}
C_0 =  N_{\rm A} k_{\rm B} \sqrt{\frac{k_{\rm B} T}{J}} \Phi^C_0\Big(\frac{g\mu_{\rm B} \mu_0 (H-H_c)}{k_{\rm B} T}\Big),\quad
C_{1} = N_{\rm A} k_{\rm B} \frac{k_{\rm B} T}{J} \Phi^C_1\Big(\frac{g\mu_{\rm B} \mu_0 (H-H_c)}{k_{\rm B} T}\Big),
\end{align}
where the two scaling functions,  $\Phi^C_0$ and $\Phi^C_1$, are straightforwardly related to $f_0$ and $f_1$ in Eq.~\eqref{FAsymptotics}. These functions and their ratio are shown in Fig.~{S3} (A) and (B), respectively. 
The relative correction $C_{1}/C_0 = \sqrt{\frac{k_{\rm B} T}{J}} \Phi^C_1/\Phi^C_0$ is most pronounced at $x \approx -2$ where $\Phi^C_1/\Phi^C_0\approx -0.4$. At $T = 0.3$ K relative corrections $C_1/C_0$ of at most $7\%$ are expected, which are too small to be identified clearly  in the experimental data.

\begin{figure}[t!]
  \centering
  \includegraphics[width=0.65\columnwidth]{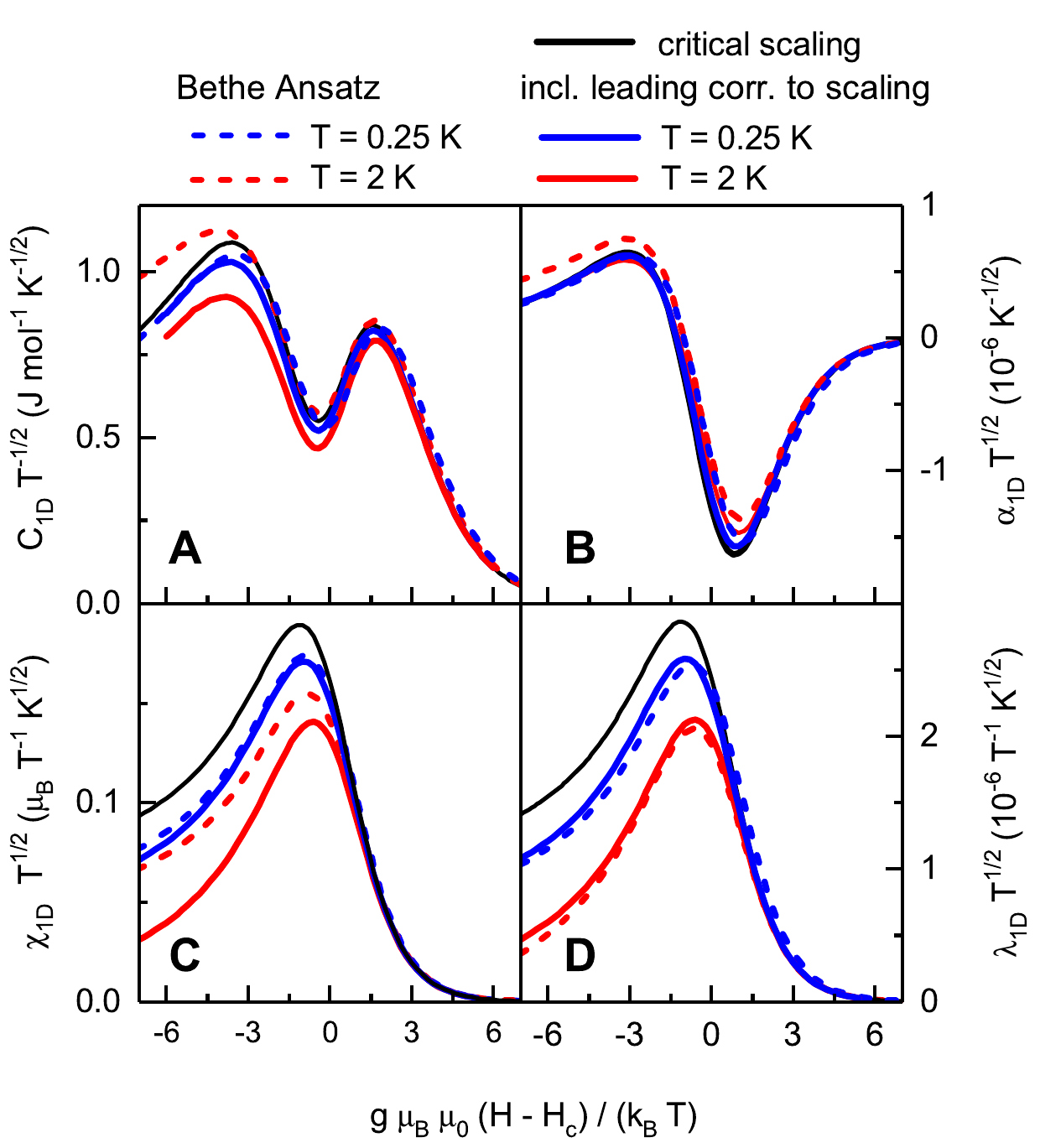}  
  \caption{{\bf Fig. S2: Theoretical prediction for the scaling of quantum critical thermodynamics in CuPzN.} The dashed lines show the exact Bethe-Ansatz result describing the experimental data as shown in Fig.~3 of the main text. The solid black line is the critical scaling asymptotics deriving from the second term in Eq.~\eqref{FAsymptotics} and the red and blue solid lines include the leading correction to scaling, i.e., also the third term in Eq.~\eqref{FAsymptotics}. The latter is sufficient for a reasonable description of the full Bethe-Ansatz result at $T = 0.25$ K but fails at $T = 2$ K, where further sub-leading corrections are important.
}  
  \label{Fig_CorrectionToScaling}
 \end{figure}

\subsubsection*{D2b. Susceptibility}

For the magnetic susceptibility, defined by $\chi = -\frac{\partial^2\mathcal{F}}{\partial(\mu_0 H)^2}$, the critical part and the leading scaling correction assume the scaling form
\begin{align}
\chi_0 = \frac{(g\mu_{\rm B})^2}{J} \sqrt{\frac{J}{k_{\rm B} T}} \Phi_0^{\chi}\Big(\frac{g\mu_{\rm B} \mu_0 (H-H_c)}{k_{\rm B} T}\Big),\quad
\chi_{1} = \frac{(g\mu_{\rm B})^2}{J} \Phi^{\chi}_{1}\Big(\frac{g\mu_{\rm B} \mu_0 (H-H_c)}{k_{\rm B} T}\Big)\, .
\end{align}
The scaling functions  $\Phi^{\chi}_{0}$ and  $\Phi^{\chi}_{1}$, resulting again from $f_0$ and $f_1$ of Eq.~\eqref{FAsymptotics}, respectively, are shown in Fig.~{S3} (C) and (D) together with their ratio $\Phi^{\chi}_{1}/\Phi^{\chi}_{0}$. This ratio is negative and monotonically decreases with decreasing argument $x$ and already exceeds $-1$ at $x\approx -3.5$. Thus, the leading corrections to scaling for $\chi$ are much larger than for the specific heat. At $T = 0.3$ K and $x\approx -5$, the relative correction $\chi_1/\chi_0$ reaches about $-21\%$, which explains the relatively large deviations of the experimental susceptibility data from the critical scaling curve $\chi_0$ for $H<H_c$, see Fig.~3 of the main text. 

\begin{figure}[t!]
  \centering
  \includegraphics[width=0.65\columnwidth]{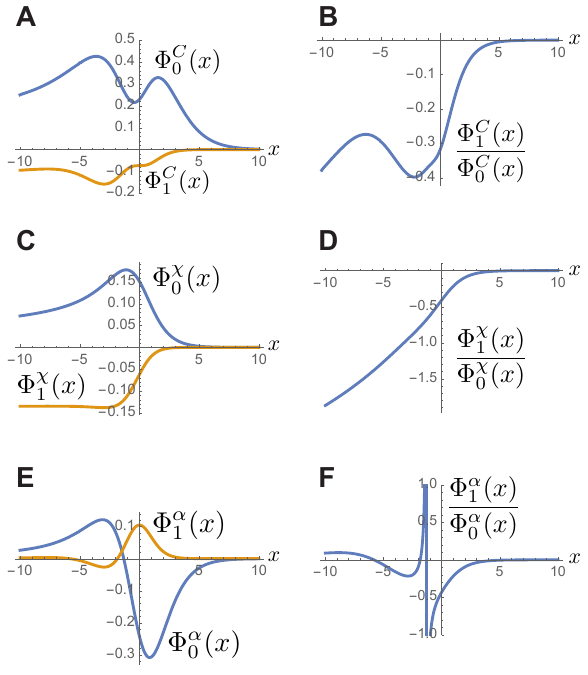}  
  \caption{{\bf Fig. S3: Deviations from critical scaling.} Scaling functions for the critical contribution, $\Phi_0$, and the leading correction to scaling, $\Phi_1$, as well as their ratio $\Phi_1/\Phi_0$ for the specific heat $C$ (A, B), the susceptibility $\chi$ (C, D), and the thermal expansion $\alpha$ (E, F). Whereas the ratio $\Phi^\chi_1/\Phi^\chi_0$ in panel (D) reaches absolute values exceeding $1.5$, the corresponding ratios in panel (B) and (C) are relatively small explaining the good scaling collapse of the specific heat and thermal expansion observed in Fig.~3 of the main text.
}  
  \label{ScalingFunctions}
 \end{figure}

\subsubsection*{D2c. Thermal expansion}

The linear thermal expansion describes the change of length $L$ upon a change of the temperature, $\alpha = \frac{1}{L}\frac{\partial L}{\partial T} = \frac{1}{V_S} \frac{\partial^2 \mathcal{F}}{\partial p\partial T}$ where $p$ is the uniaxial pressure and $V_S$ is the volume per spin. We assume that the pressure dependence arises from a magnetoelastic coupling that yields a weakly pressure dependent exchange $J(p)$. The free energy \eqref{FAsymptotics} depends on $J$ via the critical field $H_c$ in the arguments of the functions $f_0$ and $f_1$ as well as via their prefactors. 
The critical scaling contribution $\alpha_0$ to the thermal expansion as well as the leading correction to scaling $\alpha_1$ are both attributed to the pressure dependence of $H_c$, respectively. The pressure dependences of their prefactors only give rise to further, subleading corrections to scaling. Both quantities, $\alpha_0$ and $\alpha_1$, can again be cast in the scaling form
\begin{align}
\alpha_0 = \frac{k_{\rm B}}{J V_S} \frac{\partial J}{\partial p} 
\sqrt{\frac{J}{k_{\rm B} T}} \Phi_0^\alpha \Big(\frac{g\mu_{\rm B} \mu_0 (H-H_c)}{k_{\rm B} T}\Big),\quad
\alpha_1 = \frac{k_{\rm B}}{J V_S} \frac{\partial J}{\partial p} 
\Phi_1^\alpha \Big(\frac{g\mu_{\rm B} \mu_0 (H-H_c)}{k_{\rm B} T}\Big).
\end{align}
The scaling functions $\Phi_0^\alpha$ and $\Phi_1^\alpha$ again result from $f_0$ and $f_1$, respectively, and are shown together with their ratio $\Phi_1^\alpha/\Phi_0^\alpha$ in Fig.~{S3} (E) and (F). Except close to the sign change of $\Phi_0^\alpha$, the absolute value of $\Phi_1^\alpha/\Phi_0^\alpha$ is smaller than $0.2$. Thus, the leading corrections to scaling for $\alpha$ are of similar magnitude as those of the specific heat.

\subsubsection*{D2d. Magnetostriction}

Finally, we discuss the magnetostriction  $\lambda = \frac{1}{L}\frac{\partial L}{\partial (\mu_0 H)} = \frac{1}{V_S} \frac{\partial^2 \mathcal{F}}{\partial p\partial (\mu_0 H)}$. In analogy to the thermal expansion, the critical contribution $\lambda_0$ and the leading correction $\lambda_1$ derive from the pressure dependence of $H_c$ in the arguments of the functions $f_0$ and $f_1$ in Eq.~\eqref{FAsymptotics}, respectively. At this order, both quantities can be related to the corresponding ones of the susceptibility, that is: $\lambda_0 = \frac{1}{V_S} \frac{2}{g\mu_B}\frac{\partial J}{\partial p} \chi_0$ and $\lambda_1 = \frac{1}{V_S} \frac{2}{g\mu_B}\frac{\partial J}{\partial p} \chi_1$. From the discussion of the susceptibility we can thus conclude that the corrections to scaling are relatively large for the magnetostriction at $H < H_{c}$.

\end{document}